\documentclass[12pt]{article}
\usepackage{appendix}
\usepackage{amsmath}
\usepackage{dsfont}
\usepackage{epsfig}
\usepackage{graphicx}
\usepackage{color}
\usepackage{amssymb,amsmath}
\newtheorem{theo}{Theorem}

\numberwithin{equation}{section}
\newcommand{\be}{\begin{equation}}
\newcommand{\ee}{\end{equation}}
\newcommand{\ba}{\begin{array}}
\newcommand{\ea}{\end{array}}
\newcommand{\beas}{\begin{eqnarray*}}
\newcommand{\eeas}{\end{eqnarray*}}
\newcommand{\bea}{\begin{eqnarray}}
\newcommand{\eea}{\end{eqnarray}}

\newcommand{\al}{\alpha}

\newcommand{\ds}{\displaystyle}

\begin{document}

\title{Exact Solution for the Portfolio Diversification Problem Based on Maximizing the Risk Adjusted Return}
\author{Abdulnasser Hatemi-J, Mohamed Ali Hajji and Youssef El-Khatib\\
UAE University}
\maketitle
\abstract{The potential benefits of portfolio diversification have been known to investors for a long time. Markowitz (1952) suggested the seminal approach for optimizing the portfolio problem based on finding the weights as budget shares that minimize the variance of the underlying portfolio. Hatemi-J and El-Khatib (2015) suggested finding the weights that will result in maximizing the risk adjusted return of the portfolio. This approach seems to be preferred by the rational investors since it combines risk and return when the optimal budget shares are sought for. The current paper provides a general solution for this risk adjusted return problem that can be utilized for any potential number of assets that are included in the portfolio. }\\

\section{Introduction}
The potential portfolio diversification benefits that can result in reducing risk have been known to investors for a long time. However, the idea of finding the optimal weights as budget shares was pioneered by Markowitz (1952). In his seminal paper the optimal weights can be found by minimizing the variance of the portfolio subject to the budget constraint. In reality, however, it is also important for the investors to pay attention to the amount of return. An alternative approach has been suggested recently in the literature for the purpose of combining the risk and the return. Thus, the optimization problem in this setting is to find the optimal weights that will result in maximizing the risk adjusted return subject to the budget constraint according to Hatemi-J and El-Khatib (2015). However, the authors provide a closed form solution for a portfolio that consists of only two assets. The objective of the current paper is to extend their work and provide an exact solution for the risk adjusted return of a portfolio that can consist of any potential number of assets subject to the budget restriction.

The rest of the paper is organized as follows. In section 2, the optimization problem
is formulated and the general solution is derived. Section 3 provides two examples. A conclusion is given in section 4. Finally, an appendix in the end presents some pertinent derivations.

\section{The optimization problem}
Let $A_i,  i=1,2,\ldots, n$, be $n$ assets included in a portfolio. Let also  $r_i,\ i=1,2,\ldots,n,$ be the return of asset $A_i$. As in Markowitz (1952), it is assumed that the returns are normally distributed such as $r_i\sim N(\mu_i,\sigma_i)$. The variance and covariance matrix is defined as $\Omega=(\sigma_{i,j})_{1\le i,j\le n}$. Let $w_i,\ i=1,2,\ldots,n,$  be the budget shares invested in asset $A_i$.  Let $ f(w)=\sum\limits_{i=1}^n \mu_i w_i$ be the expected return of the portfolio and $g(w)=w^t\Omega w$ be the variance,  $w^t$ denotes the transpose of $w$. The risk adjusted return  of the underlying portfolio is 
\be
Q(w)=\frac{\sum\limits_{i=1}^{n} w_i \mu_i}{\sqrt{w^t\Omega w}}=\frac{f(w)}{\sqrt{g(w)}}.
\ee
  The  problem is to determine the optimal weights vector  $w=(w_i)_{1\le i\le n}$ such that the risk adjusted return, $Q(w)$, of the underlying portfolio is  maximized, subject to the budget constraint $\sum\limits_{i=1}^n w_1=1$.
 Thus, the optimization problem at hand is the following.
\be
\max_{w\in R^n}Q(w)\qquad\mbox{subject to}\qquad C(w)=\sum_{i=1}^n w_i-1=0.
\label{oppb}
\ee

 The following theorem gives  the exact optimal solution of the optimization problem (\ref{oppb}) in a compact form.
\begin{theo}\label{theorem1}
The optimal solution, $w^*$, to the maximization problem (\ref{oppb}) is given by
\be
w^*=\al-\frac{(\mu^t\beta)(\al^t\Omega\al)-(\mu^t\al)(\al^t\Omega\beta)}{(\mu^t\beta)(\al^t\Omega\beta)-(\mu^t\al)(\beta^t\Omega\beta)}\beta,
\label{wstar}
\ee
where
$$\al=\frac{\Omega^{-1}u}{\sum(\Omega^{-1} u)},\qquad \beta=\Omega^{-1}v-\left(\sum(\Omega^{-1} v) \right)\al,
$$
 with $u=(u_i)_{i=1,...,n}$ and $v=(v_i)_{i=1,...,n}$ are $n\times 1$ constant vectors
 given recursively by
\beas
&& u_{i}=a_iu_{i+1}+b_iu_{i+2},\ u_{n-1}=1,\ u_n=0,\qquad \ i=1,\ldots,n-2, \\
&& v_{i}=a_iv_{i+1}+b_iv_{i+2},\ v_{n-1}=0,\ v_n=1,\qquad \ i=1,\ldots,n-2,
\eeas
and
$$ a_i=\frac{\mu_{i+2}-\mu_{i}}{\mu_{i+2}-\mu_{i+1}},\ \ b_i=\frac{\mu_{i}-\mu_{i+1}}{\mu_{i+2}-\mu_{i+1}},\qquad \ i=1,\ldots,n-2,
$$
 For an $n\times 1$ vector $x$, the notation $\sum(\Omega^{-1}x)=\sum\limits_{i=1}^n (\Omega^{-1} x)_i$.
\end{theo}\mbox{}\\

\noindent
{\bf Proof of Theorem 1:}
Define the Lagrange function to problem (\ref{oppb}) by
\be \label{lagrangian}
L(w;\lambda)=Q(w)-\lambda C(w),\ \lambda\in R.
\ee
A solution to problem (\ref{oppb}) is solution of the system
\bea
&&\frac{\partial L}{\partial w_i}=0,\qquad i=1,\ldots,n, \label{Li}\\
&&\frac{\partial L}{\partial \lambda}=0, \label{Llambda}
\eea
Using the notation  $f_i=\frac{\partial f}{\partial w_i}$ and $g_i=\frac{\partial g}{\partial w_i}$, (\ref{Li}) and (\ref{Llambda}) reduce to
\bea
&& 2f_ig-fg_i=2g^{3/2}\lambda,\qquad  i=1,\ldots,n,
\label{seq1}\\
&&\sum_{i=1}^n w_i=1,\label{seq2}
\eea
From (\ref{seq1}), we see the right-hand side is independent of $i$. Thus, we have
$$
2f_{i+1} g - fg_{i+1}=2f_i g - fg_i,\qquad  i=1,\ldots,n-1,
$$
or
\be
\frac{g_{i+1}-g_{i}}{f_{i+1}-f_{i}}=\frac{2g}{f},\qquad  i=1,\ldots,n-1.
\label{eq2}
\ee
Again, the right-hand side of (\ref{eq2}) is independent of $i$, and we have
\be
\frac{g_{i+2}-g_{i+1}}{f_{i+2}-f_{i+1}}=\frac{g_{i+1}-g_{i}}{f_{i+1}-f_{i}},\qquad  i=1,\ldots,n-2.
\label{nm2sys}
\ee
Equation (\ref{nm2sys}) means that $w$ satisfies the homogeneous linear system
\be
(f_{i+1}-f_{i})(g_{i+2}-g_{i+1})-(f_{i+2}-f_{i+1})(g_{i+1}-g_{i})=0,\qquad i=1,\ldots,n-2.
\label{eq3}
\ee
Since $f_i=\mu_i$ and
$g_i=((\Omega+\Omega^t)w)_i=2(\Omega w)_i$,
the $(n-2)\times n$ system   (\ref{eq3})  can be written in matrix form as
\be
B\Omega w=0, \label{fsys}
\ee
with  $B$ the upper tridiagonal  $(n-2)\times n$ matrix:
$$\small
\left[\ba{ccccccc}(\mu_3-\mu_2)&(\mu_1-\mu_3)&(\mu_2-\mu_1)& 0&0&\ldots&0\\
0&(\mu_4-\mu_3)&(\mu_2-\mu_4)&(\mu_3-\mu_2)&0&\ldots&0\\
\vdots&\ddots&\ddots&\ddots&\ddots&\ddots&\vdots\\
0&0&\ldots&0&(\mu_n-\mu_{n-1})&(\mu_{n-2}-\mu_n)&(\mu_{n-1}-\mu_{n-2})
\ea
\right].
$$
The  general solution of (\ref{fsys}) can be written as
$$
w=\Omega^{-1} z,
$$
where $z=(z_i)_{i=1,...,n}$ is the general solution of the  system $Bz=0$. If we assume that $rank(B)=n-2$, i.e., $\mu_{i}\neq \mu_{i-1}, \ 3\leq i\leq n$, then $z=(z_i)_{i=1,...,n}$ is given recursively by
\be
z_{i}=a_iz_{i+1}+b_iz_{i+2},\ i=1,\ldots,n-2,
\ee
with $z_{n-1}=s,\ \ z_n=t$, $s$ and $t$ are free parameters, and
$$
a_i=\frac{\mu_{i+2}-\mu_{i}}{\mu_{i+2}-\mu_{i+1}},
\ b_i=\frac{\mu_{i}-\mu_{i+1}}{\mu_{i+2}-\mu_{i+1}},\quad \ i=1,\ldots,n-2.
$$
Since every $z_i$ is a linear combination of the free parameters $s$ and $t$, we can write $z=s u + t v$, where $u=(u_i)_{i=1,...,n}$ and $v_{i=1,...,n}$ are $n\times 1$ constant vectors
 given recursively by
\be\label{uandv}
u_{i}=a_iu_{i+1}+b_iu_{i+2},\ \ v_{i}=a_iv_{i+1}+b_iv_{i+2},\qquad \ i=1,\ldots,n-2,
\ee
with $u_{n-1}=1,\ u_n=0$ and $v_{n-1}=0,\ v_n=1$.
It follows that the general solution of (\ref{fsys}) is
$$
w=\Omega^{-1}(s u+t v)=s \Omega^{-1} u+t  \Omega^{-1} v.
$$

Imposing the condition (\ref{seq2}), $\sum\limits_{i=1}^n w_i=1$, we get
$$
\ds s=\frac{1-t \sum(\Omega^{-1} v)_i}{\sum(\Omega^{-1} u)_i}.
$$
Thus, we have $w$ as
$$
w= \al+t \beta,
$$
where
\be\label{alphabeta}
\al=\frac{\Omega^{-1}u}{\sum(\Omega^{-1} u)_i},\qquad
 \beta=\Omega^{-1}v-\left(\sum(\Omega^{-1} v)_i \right)\al.
\ee
 From the above derivations, we conclude that the optimal solution $w^*$ is of the form
$$
w^*=\al+t^*\beta
$$
for some optimal value $t^*$, with $\al$ and $\beta$ as defined above. The optimal value $t^*$ is the one that maximizes the risk adjust return $Q(\al+t \beta)$, i.e., $\frac{\partial Q(\al+t \beta)}{\partial t}=0$. Derivations (see  Apprendix A) reveal that
\be
t^*=-\frac{(\mu^t\beta)(\al^t\Omega\al)-(\mu^t\al)(\al^t\Omega\beta)}
{(\mu^t\beta)(\al^t\Omega\beta)-(\mu^t\al)(\beta^t\Omega\beta)}.
\label{tstar}
\ee
Therefore, the optimal weights vector is given 
\be
w^*=\al-\frac{(\mu^t\beta)(\al^t\Omega\al)-(\mu^t\al)(\al^t\Omega\beta)}
{(\mu^t\beta)(\al^t\Omega\beta)-(\mu^t\al)(\beta^t\Omega\beta)}\beta,
\label{ws}
\ee
which proves the Theorem \ref{theorem1}.

\section{Examples}
In this section, we validate the formula (\ref{ws}) by applying it to the case of a portfolio consisting of two assets, and extend it to the 3 dimensional case.\\

\noindent
{\bf Example 1.}  In the case $n=2$, we have $\mu=[\mu_1\ \mu_2]^t$, $\Omega=\left[\ba{cc}\sigma_{11}& \sigma_{12}\\\sigma_{21}&\sigma_{22}\ea\right],\ \sigma_{12}=\sigma_{21}$. From equation (\ref{uandv}), we have $u=[u_1\ u_2]^{t}=[1\ 0]^t$, $v=[v_1\ v_2]^{t}=[0\ 1]^t$ and from equation (\ref{alphabeta}), we have 
$$ \al=\left(\frac{\sigma _{2,2}}{\sigma _{2,2}-\sigma _{1,2}},\frac{-\sigma _{1,2}}{\sigma
   _{2,2}-\sigma _{1,2}}\right),\qquad
  \beta= \left(\frac{-1}{\sigma _{2,2}-\sigma _{1,2}},\frac{1}{\sigma _{2,2}-\sigma
   _{1,2}}\right).  $$
Substituting these into (\ref{ws}), we obtain
\beas
&& w_1=\frac{\mu_1 \sigma_{2,2}-\mu_2\sigma_{1,2}}{\mu_1(\sigma_{2,2}-\sigma_{1,2})+\mu_2(\sigma_{1,1}-\sigma_{1,2})},\\
&&w_2=\frac{\mu_2 \sigma_{11}-\mu_1 \sigma_{12}}{\mu_1(\sigma_{2,2}-\sigma_{1,2})+\mu_2(\sigma_{1,1}-\sigma_{1,2})},
\eeas
as found in Hatemi-J and   El-Khatib.\\

\noindent
{\bf Example 2.}   In the case $n=3$, we have $\mu=[\mu_1\ \mu_2\ \mu_3]^t$, $\Omega=\left[\ba{ccc}\sigma_{11}& \sigma_{12}& \sigma_{13}\\\sigma_{21}&\sigma_{22}&\sigma_{23}\\
\sigma_{31}& \sigma_{32}& \sigma_{33}\ea\right]$, where $\sigma_{2,1}=\sigma_{1,2}, \ \sigma_{3,1}=\sigma_{1,3}$, and $\sigma_{3,2}=\sigma_{2,3}$. From equation (\ref{uandv}), the vectors $u$ and $v$, from $u=[u_1\ u_2\ u_3]^{t}=\left\{\frac{\mu _3-\mu _1}{\mu _3-\mu _2},1,0\right\}^t$, $v=[v_1\ v_2\ v_3]^{t}=\left\{\frac{\mu _1-\mu _2}{\mu _3-\mu _2},0,1\right\}^t$. Similarly, we obtain the $\alpha$ and $\beta$ vectors and when  
Substituted  into (\ref{ws}), we obtain
\beas
&& w_1= 
\frac{\mu _3 \left(\sigma _{1,3} \sigma _{2,2}-\sigma _{1,2} \sigma _{2,3}\right)+\mu _2
   \left(\sigma _{1,2} \sigma _{3,3}-\sigma _{1,3} \sigma _{2,3}\right)+\mu _1
   \left(\sigma _{2,3}^2-\sigma _{2,2} \sigma _{3,3}\right)}{\Delta},
\\
&&w_2=\frac{\mu _3 \left(\sigma _{1,1} \sigma _{2,3}-\sigma _{1,2} \sigma _{1,3}\right)+\mu _2
   \left(\sigma _{1,3}^2-\sigma _{1,1} \sigma _{3,3}\right)+\mu _1 \left(\sigma _{1,2}
   \sigma _{3,3}-\sigma _{1,3} \sigma _{2,3}\right)}{\Delta},
\\
&&w_3=\frac{\mu _3 \left(\sigma _{1,2}^2-\sigma _{1,1} \sigma _{2,2}\right)+\mu _2 \left(\sigma
   _{1,1} \sigma _{2,3}-\sigma _{1,2} \sigma _{1,3}\right)+\mu _1 \left(\sigma _{1,3}
   \sigma _{2,2}-\sigma _{1,2} \sigma _{2,3}\right)}{\Delta},
\eeas
where
\beas
\Delta&=&\mu _3 \left(\sigma
   _{1,2}^2-\left(\sigma _{1,3}+\sigma _{2,3}\right) \sigma _{1,2}+\left(\sigma
   _{1,3}-\sigma _{1,1}\right) \sigma _{2,2}+\sigma _{1,1} \sigma _{2,3}\right)\\
   && +\mu _2
   \left(-\sigma _{1,3} \left(\sigma _{1,2}-\sigma _{1,3}+\sigma _{2,3}\right)+\sigma
   _{1,1} \left(\sigma _{2,3}-\sigma _{3,3}\right)+\sigma _{1,2} \sigma _{3,3}\right)\\
   &&+\mu_1 \left(\sigma _{1,3} \left(\sigma _{2,2}-\sigma _{2,3}\right)+\sigma _{2,3}
   \left(\sigma _{2,3}-\sigma _{1,2}\right)+\left(\sigma _{1,2}-\sigma _{2,2}\right)
   \sigma _{3,3}\right)
   \eeas

\section{Conclusion}
Investors regularly make use of portfolios in order to remove or at least reduce the unsystematic risk of their investments. A critical issue within this context is finding the optimal weights as budget shares. The seminal paper of Markowitz (1952) suggests finding the optimal  weights that result in minimum variance as a measure of risk of the underlying portfolio. This classical approach might lead to selecting a portfolio that would be the safest but not necessarily the one that would give the highest amount of return per unit of risk. Hatemi-J and El-Khatib (2015) suggest finding the optimal weights that combine risk and return based on the maximization of the risk adjusted return. They managed to provide solution for the $2\times 2$ case. However, in financial markets investors are interested usually in portfolios that will include more than two assets because the diversification benefit is an increasing function of the number of assets included in that portfolio.  The current paper provides closed form solutions for a portfolio that is based on the risk adjusted return of dimension  $n\times n$, where $n$ represents the number of assets.

In the future, applications can be provided to show that the risk adjusted return of a portfolio based on our suggested approach has higher value compared to the risk adjusted return of other alternative portfolios. It should be also mentioned that this method can be extended to the financial markets in which short selling is not possible, that is, all weights must be positive. Other extensions could be allowing for higher moments in the data generating process of the underlying assets such as the third and the fourth moments.

\appendix
\section{Appendix}
In this appendix we  present the derivation of the formula for $t^*$ given in (\ref{tstar}). Let $\alpha$ and $\beta$ be two constant $n\times 1$ column vectors. Let 
$$h(t)=Q(\alpha+t\beta)=\frac{f(\alpha+t\beta)}{\sqrt{g(\alpha+t\beta)}}.
$$
The optimal $t^*$ is the value of $t$ for which $h'(t)=0$, where 
$$
h'(t)= \frac{\partial Q}{\partial t}=\frac{2f_t g -f g_t}{2g^{3/2}}.
$$
Then $t^*$ is the value of $t$ for which 
\be
2f_t g -f g_t=0.
\label{numeq0}
\ee
We have
\beas
f(\alpha+t\beta)&=&\sum\limits_{i=1}^n\mu_i(\al_i+t\beta_i)\Longrightarrow f_t=\sum\limits_{i=1}^n \mu_i\beta_i\\
 g(\alpha+t\beta)&=& \sum_{i,j}\sigma_{i,j} \al_i\al_j+\sum_{i,j}\sigma_{i,j} (\al_i\beta_j+\al_j\beta_i)t+\sum_{i,j}\sigma_{i,j} (\beta_i\beta_j)t^2\\
\Longrightarrow g_t&=&\sum_{i,j}\sigma_{i,j} (\al_i\beta_j+\al_j\beta_i)+2\sum_{i,j}\sigma_{i,j} (\beta_i\beta_j)t
\eeas
Then (\ref{numeq0}) becomes
\beas
2f_t g -f g_t&=&2\left(\sum\limits_{i=1}^{n}\mu_i\beta_i\right)\left(\sum_{i,j=1}^n \sigma_{i,j} \al_i\al_j+\sum_{i,j=1}^n \sigma_{i,j} (\al_i\beta_j+\al_j\beta_i)t+\sum_{i,j=1}^n\sigma_{i,j} (\beta_i\beta_j)t^2\right)\\
&& -\left(\sum\limits_{i=1}^n\mu_i(\al_i+t\beta_i)\right)\left(\sum_{i,j=1}^n \sigma_{i,j} (\al_i\beta_j+\al_j\beta_i)+2\sum_{i,j=1}^n\sigma_{i,j} (\beta_i\beta_j)t\right)\\
&=&\ \ \ \left[2\left(\sum\limits_{i=1}^n\mu_i\beta_i\right)\left(\sum_{i,j=1}^n\sigma_{i,j} \al_i\al_j\right)
-\left(\sum_{i=1}^n \mu_i\al_i\right)\left(\sum_{i,j=1}^n\sigma_{i,j}(\al_i\beta_j+\al_j\beta_i)\right)\right]\\
&&+t\bigg[2\left(\sum_{i=1}^n\mu_i\beta_i\right)\left(\sum_{i,j=1}^n\sigma_{i,j}(\al_i\beta_j+\al_j\beta_i)\right)
-\left(\sum_{i=1}^n\mu_i\beta_i\right)\left(\sum_{i,j=1}^n\sigma_{i,j}(\al_i\beta_j+\al_j\beta_i)\right)\\
&&\qquad-2\left(\sum_{i=1}^n\mu_i\al_i\right)\left(\sum_{i,j=1}^n\sigma_{i,j}(\beta_i\beta_j)\right)
\bigg]\\
&&+t^2\bigg[\underbrace{2\left(\sum_{i=1}^n\mu_i\beta_i\right)\left(\sum_{i,j=1}^n\sigma_{i,j}(\beta_i\beta_j)\right)
-2\left(\sum_{i=1}^n\mu_i\beta_i\right)\left(\sum_{i,j=1}^n\sigma_{i,j}(\beta_i\beta_j)\right)}_{=0}
\bigg]\\
&=&2(\mu^t\beta)(\al^t\Omega\al)-(\mu^t\al)(\al^t\Omega\beta+\beta^t\Omega\al)+t\left[(\mu^t\beta)(\al^t\Omega\beta+\beta^t\Omega\al)-2(\mu^t\al)(\beta^t\Omega\beta)\right]
\eeas
It follows that $2f_t g -f g_t=0$ for 
$$
t^*=-\frac{2(\mu^t\beta)(\al^t\Omega\al)-(\mu^t\al)(\al^t\Omega\beta+\beta^t\Omega\al)}{(\mu^t\beta)(\al^t\Omega\beta+\beta^t\Omega\al)-2(\mu^t\al)(\beta^t\Omega\beta)}
$$
Since $\Omega$ is symmetric, we have $\beta^t\Omega\al=\al^t\Omega\beta$ and $t^*$ reduces to 
$$
t^*=-\frac{(\mu^t\beta)(\al^t\Omega\al)-(\mu^t\al)(\al^t\Omega\beta)}{(\mu^t\beta)(\al^t\Omega\beta)-(\mu^t\al)(\beta^t\Omega\beta)},
$$
as in (\ref{tstar}).

\end{document}